\documentclass[twocolumn]{aastex63}
\usepackage{tabularx}

\newcommand{\ith}{\ensuremath{^{\rm th}}}

\newcommand{\logg}{\ensuremath{\log g}}

\submitjournal{\textit{The Astronomical Journal}}

\graphicspath{{./}{figures/}}
\newcommand{\eg}{{\it e.g.}}

\shorttitle{\texttt{Astraea}: Predicting Long Rotation Periods with 27-Day Light Curves}
\shortauthors{Lu et al.}

\begin{document}

\title{\texttt{Astraea}: Predicting Long Rotation Periods with 27-Day Light Curves}

\correspondingauthor{Yuxi(Lucy) Lu}
\email{lucylulu12311@gmail.com}

% Affiliations
\newcommand{\amnh}{American Museum of Natural History, Central Park West, Manhattan, NY, USA}
\newcommand{\cca}{Center for Computational Astrophysics, Flatiron Institute, 162 5\ith\ Avenue, Manhattan, NY, USA}
\newcommand{\columbia}{Department of Astronomy, Columbia University, 550 West 120\ith\ Street, New York, NY, USA}

% Author List
\author[0000-0003-4769-3273]{Yuxi(Lucy) Lu}
\affiliation{\columbia}
\affiliation{\amnh}

\author[0000-0003-4540-5661]{Ruth Angus}
\affiliation{\amnh}
\affiliation{\cca}
\affiliation{\columbia}

\author[0000-0001-7077-3664]{Marcel A.~Ag\"{u}eros}
\affiliation{\columbia}

\author{Kirsten Blancato}
\affiliation{\columbia}

\author{Melissa Ness}
\affiliation{\columbia}

\author[0000-0003-2144-4885]{Danielle Rowland}
\affiliation{\amnh}

\author[0000-0002-2792-134X]{Jason L.~Curtis}
\affiliation{\amnh}

\author[0000-0003-4976-9980]{Sam Grunblatt}
\affiliation{\amnh}
\affiliation{\cca}

% Start a new line for every sentence -- makes it easier to comment out blocks!
\begin{abstract}
The rotation periods of planet-hosting stars can be used for modeling and mitigating the impact of magnetic activity in radial velocity measurements, and can help constrain the high-energy flux environment and space weather of planetary systems.
Millions of stars and thousands of planet hosts are observed with the Transiting Exoplanet Survey Satellite (TESS).
However, most will only be observed for 27 contiguous days in a year, making it difficult to measure rotation periods with traditional methods.
This is especially problematic for field M dwarfs, which are ideal candidates for exoplanet searches, but which tend to have periods in excess of the 27-day observing baseline.
We present a new tool, \texttt{Astraea}, for predicting long rotation periods from short-duration light curves combined with stellar parameters from {\it Gaia} DR2.
Using \texttt{Astraea}, we can predict the rotation periods from Kepler 4-year light curves with 13\% uncertainty overall (and a 9\% uncertainty for periods $>$30 days).
By training on 27-day Kepler light curve segments, \texttt{Astraea} can predict rotation periods up to 150 days with 9\% uncertainty (5\% for periods $>$30 days).
After training this tool on these 27-day Kepler light curve segments, we applied \texttt{Astraea} to real TESS data.
For the 195 stars that were observed by both Kepler and TESS, we were able to predict the rotation periods with 55\% uncertainty despite the wild differences in systematics.

\end{abstract}


%% Barnes2007,Chaplin2014 (asteroseismology)
\section{Introduction} \label{sec:intro}
The rotation period of a star is one of the most direct observables one can measure.
It is closely linked with its physical parameters such as magnetic activity, surface gravity and even stellar age \citep[e.g.][]{Skumanich1972, Barnes2007, McQuillan2014, Davenport2019,vanSaders2013}.
Rotation periods can be used to age-date stars via ``gyrochronology'' \citep[e.g.][]{Barnes2003, Barnes2007}, study the internal structures of stars, learn about stellar magnetic fields, and improve the precision of exoplanet detection.

In the field of exoplanet detection, additional astrophysical signals tied to stellar rotation can often complicate the process.
For example, the effects of stellar magnetism in rotating stars can negatively affect exoplanet detection or characterization using radial velocity (RV) measurements.
Dark spots and bright plages on the surface of a rotating star can alter the profiles of spectral absorption lines and introduce signals into RV time series.
These effects are normally weak and can be treated as background noise in pipelines for discovering exoplanets.
However, in the case of a planet orbiting an active star, the RV signal from the planet can be embedded within that from the host star and thus, making planet signal extraction difficult \citep[e.g.][]{Hillenbrand2015,Haywood2014,Rajpaul2015}.
Modeling both the stellar activity from the host star and the orbital parameters of the planet simultaneously is essential in these scenarios.
Furthermore, knowing the rotation period of the star can assist in improving the model \cite[e.g.][]{Grunblatt2015}.
%These effects can modeled if the rotation period of the star is known, but stars with rotation periods that are similar to the orbital periods of their planets, or a harmonic, present challenging follow-up targets.
%If the orbital period of the planet differs from the rotational period of the star, stellar and planetary RV signals can often be modeled simultaneously without needing to know the stellar rotation period in advance \textcolor{red}{(cite examples)}.
%However, knowing the stellar rotation period will always improve models of the stellar activity and, in cases where rotational and orbital periods of star and planet are similar, it is essential that the rotation period of the star in known.

M dwarfs are also the most suitable host stars for finding rocky planets in the habitable zone since these stars are small (so the transit signal is larger) and dim (so the habitable zone is closer).
This means the transit and radial velocity signals from small planets orbiting an M dwarf are stronger compared to those orbiting more massive, large host stars.
However, the rotation periods of M dwarfs are often longer than the typical observing window of {\it TESS} (27.4 days), so non-standard methods must be used to measure their rotation periods.

The most common tools used to measure rotation periods are Lomb--Scargle periodograms \citep[e.g.][]{Reinhold2015}, Auto-Correlation Functions (ACFs) \citep[e.g.][]{McQuillan2014} and Gaussian processes \citep[e.g.][]{Angus2018,ForemanMackey2017}.
These methods typically require the observed light curve to contain continuous data for more than one rotation period of the star in order to get an accurate estimate.
Long rotation periods can be measured precisely for stars observed by {\it Kepler} \citep{Borucki2010} that show periodic signals.
However, long rotation periods for stars observed by {\it TESS}, especially those with only 27 days of observations per year \citep[fall in this category;][]{TESS}, are extremely hard to measure directly.
Even more challenging, low-mass stars (e.g. M dwarf stars) usually have long rotation periods \citep[$>$ 25-30 days;][]{McQuillan2014}.
Because of this, traditional methods will not be able to provide accurate or precise rotation period measurements for most M dwarfs using {\it TESS} single-sector light curves.

As we know from empirical gyrochronology studies \citep[e.g.][]{Barnes2003, Barnes2007,VanSaders2016}, the rotation period of a star is mainly determined by its age and color.
Therefore, if it were possible to measure the ages of stars precisely, we could accurately predict their rotation periods.
However, the relation between stellar rotation, age and color could breakdown at a high Rossby number (rotation period divided by the local convective turnover time).
\cite{VanSaders2016} pointed out magnetic weakening may cause stalling in stellar spin-down for Rossby number greater than $\sim$ 2, and will cause gyrochronology relations to break down approximately half way through the stars' main-sequence lifetime. 
This effect means we may not be able to predict rotation periods for stars that have already gone through half of their main-sequence lifetime.
Fortunately, this effect is not significant for the catalogs we used in this study from \cite{McQuillan2014,Santos2019,Garcia2014}.

However, the ages of stars, especially low-mass dwarfs, are extremely difficult to measure \citep[see e.g. ][for a review of stellar ages]{soderblom2010}.
Fortunately, there are many easily observable, indirect age proxies that can be used in lieu of directly measured ages (the relations are very complex and thus ages are very hard to predict for main-sequence stars).
For example, stellar velocity, radius, and surface gravity are all related, albeit weakly, to stellar age.
Therefore, we expect to be able to extract information about rotation periods from these stellar properties.
However, since the relationship between these properties and rotation period is “weak” and potentially non-linear, a machine learning approach can be used to combine these properties with observables such as color, surface temperature or mass information, to accurately predict stellar rotation periods.

In addition, there are some other potential indirect age proxies we can measure:

{\it Flicker ---} the brightness variation on timescales of 8 hours and less caused by convection-driven fluctuations on the stellar surface \citep{Bastien2013}.
By comparing flicker with asteroseismic \logg, \cite{Bastien2013} concluded that \logg\ can be estimated from flicker with $\sim$ 0.1 dex uncertainty.
If we are able to measure flickers for main-sequence stars, these measurements should be able to provide information about the surface gravity, which decreases as a star ages. 
As a result, it is possible to predict rotation periods by combining flicker with other stellar properties.
One of the advantages of using this method is that flicker occurs on very short timescales.
Therefore, we can extract granulation signals from light curves that are only 27-days long.
However, flicker can be hard to measure in the light curves of M dwarfs due to the granulation signal being weak. 

{\it Flaring activities ---} both a star's flare energy and frequency of young, active stars are associated with their ages and rotation periods \citep{Davenport2019}.
Since low-mass stars have deeper convective envelopes that are associated with stronger magnetic fields, flares are more commonly detected in these stars \citep{Ilin2019}.
Therefore, flare rates could potentially be an indicator of the rotation periods of M dwarfs.
However, one major limitation is that for inactive stars, which are typically older and have longer rotation periods, the rate of flaring is often too low to be detected within the short 27-day light curves of \textit{TESS}.

{\it Stellar kinematics ---} The kinematic properties of a star is shown to be related to the age of a star.
For example, the vertical velocity dispersion of stars increase over time at a rate that can be quantified with an age-velocity dispersion relation (AVR).
\cite{Stromberg1946} and \cite{Spitzer1951} first discovered older stars have higher vertical velocity dispersion and this relation has been confirmed by further observations \citep[\eg][]{Nordstrom2004,Holmberg2007,Holmberg2009,Aumer2009,Yu2018,Ting2019}.
Two possible theories can explain these observations.
One such theory is that all stars formed kinematically `cool' and as the Milky Way evolved, older stars were scattered to higher galactic latitudes by the giant molecular clouds and spiral arms.
Therefore, these older stars have a higher velocity dispersion \citep[\eg][]{sellwood2014, Lacey1984, Barbanis1967, Sellwood1984}.
Another theory is that these older stars were born kinematically `hot' in the first place \citep[\eg][]{Bird2013}.
Since the age and rotation period of a main-sequence star are correlated, the velocity dispersion, or other kinematic information (e.g. vertical velocity, galactic latitude, etc.), could also be useful in determining rotation periods of stars. 

Although there are many stellar properties closely tied to the rotation period of a star, it is hard to model the relations between stellar rotation and other physical properties.
Low order polynomial fits are often used to interpolate these relationships, but it is clear that the correlations are not simple.
Machine learning (ML) algorithms are particularly good at modeling complex, non-linear relations.
A ML model is normally trained on a large training data set for it to learn the complex relations between features and labels in the data.
In this project, the features are the stellar properties (e.g. surface temperature, radius, color, etc.) and the label is the rotation period.
After being trained, the ML model will be able to predict labels from features at a very fast speed. 
In addition, the same ML models can be adapted to different missions fairly straightforwardly by using the right training data.
As a result, ML algorithms are likely to become more popular as astronomers march into the big data era.
In particular, current and future missions observing stars, such as {\it Kepler} \citep{Borucki2010}, {\it Gaia} \citep{Prusti2016, Brown2018}, {\it TESS} \citep{TESS}, Vera C. Rubin Observatory \citep{LSST} and {\it Planetary Transits and Oscillations of stars (PLATO)} \citep{PLATO} will require rapid data processing algorithms to accommodate the large data flow.
It is essential to analyze data quickly and efficiently in order to maximize the information usage of these missions.
Another benefit of using data-driven ML algorithms is that we can get insight on the data set itself. 
For example, a trained ML model can identify interesting anomalies or outliers in the data.
We will describe briefly how the ML model we trained could potentially be used as a binary identifier in section \ref{sec:disc}.

We use a particularly well-studied machine learning approach of Random Forest (RF) \citep{Breiman2001} to predict the rotation periods for stars in the {\it TESS} 27-day observing fields, based on their stellar properties (Table~\ref{tab:feature} shows the list of properties used to predict rotation periods). 
RF is a machine learning algorithm that combines multiple decision trees to prevent over-fitting, and a suitable algorithm to learn complex non-linear relations between different stellar properties.
Decision trees use multiple parameters (e.g. effective temperature, radius, luminosity, etc.), which are often called {\it `features'}, to split the data into different subsets (where the data split is called a {\it `node'}.) and predict the {\it `label'} (e.g. rotation period).
A RF algorithm trains a number of decision trees on different subsets of the data and predicts the label by averaging the resulting predictions from each decision tree.
This machine learning approach has huge potential to automate the delivery of rotation period from observation data.
RF, compared to neural networks or deep learning, is relatively easier to interpret since the input features are selected by the user and the user can calculate the feature importance and gain insight into the data itself.
This method can be used to capture and effectively model the relationships between stellar rotation, stellar age and stellar parameters including temperature, radius, and surface gravity.
RFs are already used in astronomy, both in classification and regression problems.
For example, \cite{Richards2011} classified variable stars with sparse and noisy time-series data with a $\sim$ 20\% error, and \cite{Miller2015} inferred fundamental stellar parameters for $\sim$ 54,000 known variables with a RF regressor.

In this paper, we exploit the relationships between rotation periods and other fundamental stellar parameters, which occur as a result of stellar evolution.
We predict rotation periods without requiring long time-series observations using a RF algorithm.
The features we used to train the model and their origins are described in section \ref{sec:datamethod}.
We first classify stars to determine whether their rotation periods are measurable,  and then use a RF regressor to predict the rotation periods of those classified as `measurable'.
The details of how we trained and optimized the models are described in section \ref{sec:Optimize}, the testing results for {\it Kepler} and {\it TESS} stars are described in section \ref{sec:results}.
Limitations and reasons we are able to predict long rotation periods from short-duration light curves are discussed in section \ref{sec:disc}. 

\section{Data \& Methods}\label{sec:datamethod}
In order to train and test a ML model, we need both a {\it training data set} (section \ref{sec:KeplerData}) and a {\it testing data set} (section \ref{sec:TESSData}).
The purpose of a training data set is to train the model to learn the complex relations between a number of ``{\it features}'' (stellar properties) (section \ref{sec:feature}) and the ``{\it label}'' (rotation period).
The purpose of a testing set is to have a number of stars that are not from the training set to validate the trained model. 

After constructing a reasonable training and testing set, we selected the useful stellar properties that are important to predict the rotation period in section \ref{sec:feature}.
One of the features we focused on is the variability of the light curve, which normally is the flux variation (range or standard deviation) averaged over one or multiple rotation period(s). 
Since we do not have information on the rotation periods, we will discuss how we can use the flux variation over the entire observing period to approximate the variability of the light curve (see section \ref{sec:feature}).

The training process for the RF classifier and the RF regressor is described in section \ref{sec:RF}.
A classifier is used to identify group(s) of data that are similar.
A regressor is used to model the complex relationships between features and labels in order to predict new labels from new features (the simplest regressor is a linear regressor).

\begin{figure*}
    \centering
    \includegraphics[width=\textwidth]{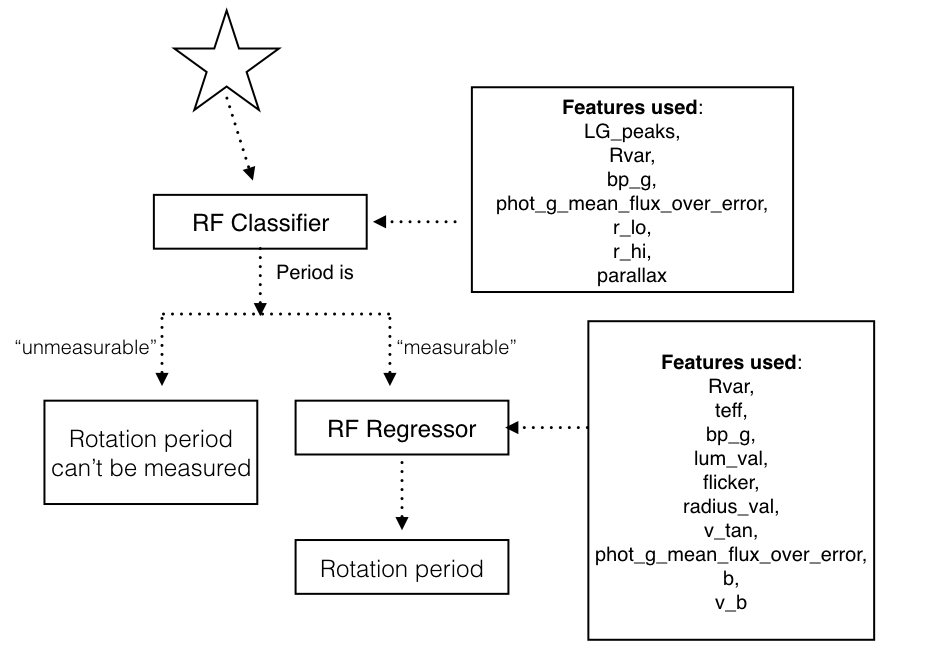}
    \caption{Pipeline of \texttt{Astraea}, our open-source software developed in this project.
    Features (stellar properties and light curve statistics) of the star are first passed through the RF classifier to identify whether the period is ``measureable'' or not.
    If the period is ``measureable'', the features will then be passed through the RF regressor to predict its stellar rotation.
    The feature descriptions are provided in Table~\ref{tab:feature}.}
    \label{fig:flow}
\end{figure*}

By combining a classifier and a regressor, we are able to classify whether a star has a ``measureable'' stellar rotation period or not, and predict its period if the period is ``measurable.''
Figure~\ref{fig:flow} shows the pipeline of \texttt{ Astraea},\footnote{Available at: https://astraea.readthedocs.io/en/latest/} the RF package (classifier + regressor) used to predict rotation period from stellar parameters.
Details of how it is built will be described in this section.

\subsection{Kepler Training Set} \label{sec:KeplerData}
We selected our training data from the {\it Kepler} field, since there already exist catalogs for the rotation periods of these stars and long rotation periods measured from 4-year {\it Kepler} light curves are more reliable.
The majority of rotation periods we used to train our models were from \cite{McQuillan2014}.
They analyzed 133,030 main-sequence {\it Kepler} targets and measured rotation periods (between 0.2 and 70 days) for 34,030 stars by using an automated ACF-based method.
ACF-based method has its advantages over Fourier based or Lomb--Scargle periodogram because the rotation period signals in the light curves are not purely sinusoidal nor strictly periodic.

%The light curves used in this project were downloaded from MAST (\dataset[doi:10.17909/T9RP4V]{https://dx.doi.org/10.17909/T9RP4V}).

\begin{figure}
    \centering
    \includegraphics[width=0.45\textwidth]{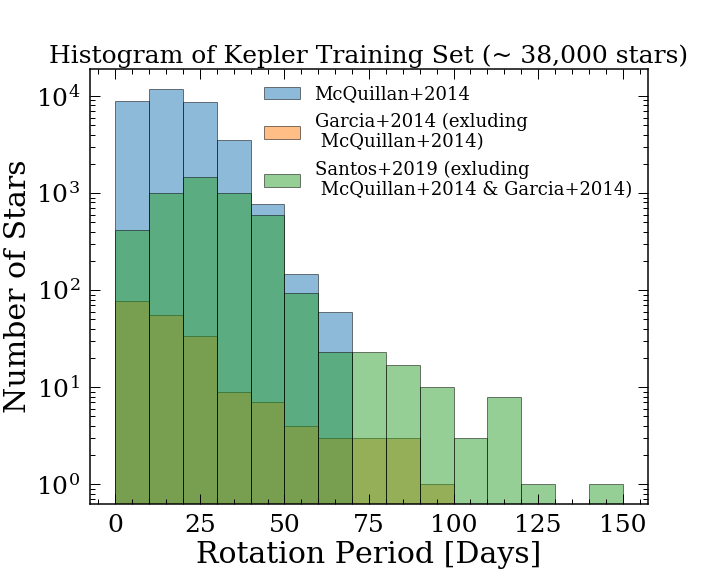}
    \caption{Histogram of the rotation periods in the {\it Kepler} training set. 
    34,030 stars from \cite{Garcia2014}, and 4,637 stars from \cite{Santos2019} and \cite{Garcia2014}.
    Supplementing the main set with these later catalogs increased the number of long period rotators, including 70 stars with periods longer than $>$ 70 days.
    Note that the $y$-axis is in log scale.}
    \label{fig:Trainingset}
\end{figure}

We utilized all the 133,030 main-sequence stars analyzed in \cite{McQuillan2014} to train a model to determine whether the rotation period for a star can be obtained.
Since our main goal was to predict long rotation periods from short-duration light curves, in addition to the 34,030 stars with rotation period measurements from \cite{McQuillan2014}, we also added 4,637 stars that have rotation periods up to $\sim$ 150 days from \cite{Santos2019} and \cite{Garcia2014}, in which they used a combination of wavelet analysis and the ACF to measure the periods.
Within these added rotation periods, 70 of them have rotation period $>$ 70 days.
This provided us with $\sim$ 38,000 {\it Kepler} stars.
Figure~\ref{fig:Trainingset} shows a histogram of the rotation periods in our training set. 

We split the data into the training data set and a validation data set so we can train our model on the training set and tune our trained model on the validation set. 
The difference between a validation data set and a testing set is subtle, but the validation data is typically used to tune the hyper-parameters (parameters relate to the ML model, see section \ref{sec:Optimize}) and the testing set is used to test the optimized model. 
Validation and testing set are both important because although the validation set can be used to optimized the model, in order to make sure the ML model is not over-fitting the validation data, a testing set is needed to test the final optimized model.
The training set is composed of 80\% random selection of stars from \cite{McQuillan2014} and the 4,637 stars from \cite{Garcia2014} and \cite{Santos2019}.
The validation set is the remaining 20\% stars. 

\subsection{{\it TESS} Test Set} \label{sec:TESSData}
After cross-matching with the {\it TESS} light curve data base hosted by the Mikulski Achieve Space Telescopes (MAST)\footnote{https://archive.stsci.edu/access-mast-data}, we were able to find 205 {\it Kepler} targets with {\it TESS} 2-minute cadence PDCSAP light curves.
We excluded 10 stars from the equal-mass binary sequence by performing a magnitude cut on the color-magnitude diagram (CMD).
A star in a unresolved close binary system with an equal masses companion will not affect its color but double its luminosity due to the starlight from its equal-mass companion. 
As a result, these stars will lay on the CMD above the main sequence stars and form a ``binary sequence.''
To exclude these stars, we first fitted a 6\ith-order polynomial to the McQuillan data sample in CMD and shifted the function up by $\sim$ 0.27 dex in absolute $G$ magnitude and excluded any stars lying above the shifted polynomial function.
After the cut, we were able to obtain a testing set of 195 stars.

\subsection{Feature Selection} \label{sec:feature}
\textit{Measuring Variability ---}
The brightness variation due to magnetic activity on the surface of a star has been shown to correlate with stellar activity, and therefore should be related to the rotation period \citep[\eg][]{McQuillan2014,Santos2019,Pizzolato2003,Hartman2011,Davies2015}.
However, brightness variation from a light curve includes more than the magnetic activity from the surface. 
Granulation, instrumental noise, p-mode oscillations, etc can also modulate a light curve. 
As a result, ideally, we would measure the light curve variability taking into account the stellar rotation period.
Two popular measurements are average amplitude of variability within one period and the standard deviation of a sub-series of length 5 $\times$ the rotation period, and these can be parameterized by $R_{per}$ or $S_{ph}$ respectively.
These two variables take into account the rotation period of a star and are shown to be closely correlated with the magnetic activity and rotation period of a star.
However, in order to measure these quantities accurately, the stars would have had gone through more than one full revolution in the observation window.
For stars observed by {\it TESS}, most slow rotators have not gone through even one full revolution within the 27-day observing period.
Therefore, it is almost impossible to get accurate measurements for either quantity, especially for the slow rotators.

\begin{figure*}
    \centering
    \includegraphics[width=\textwidth]{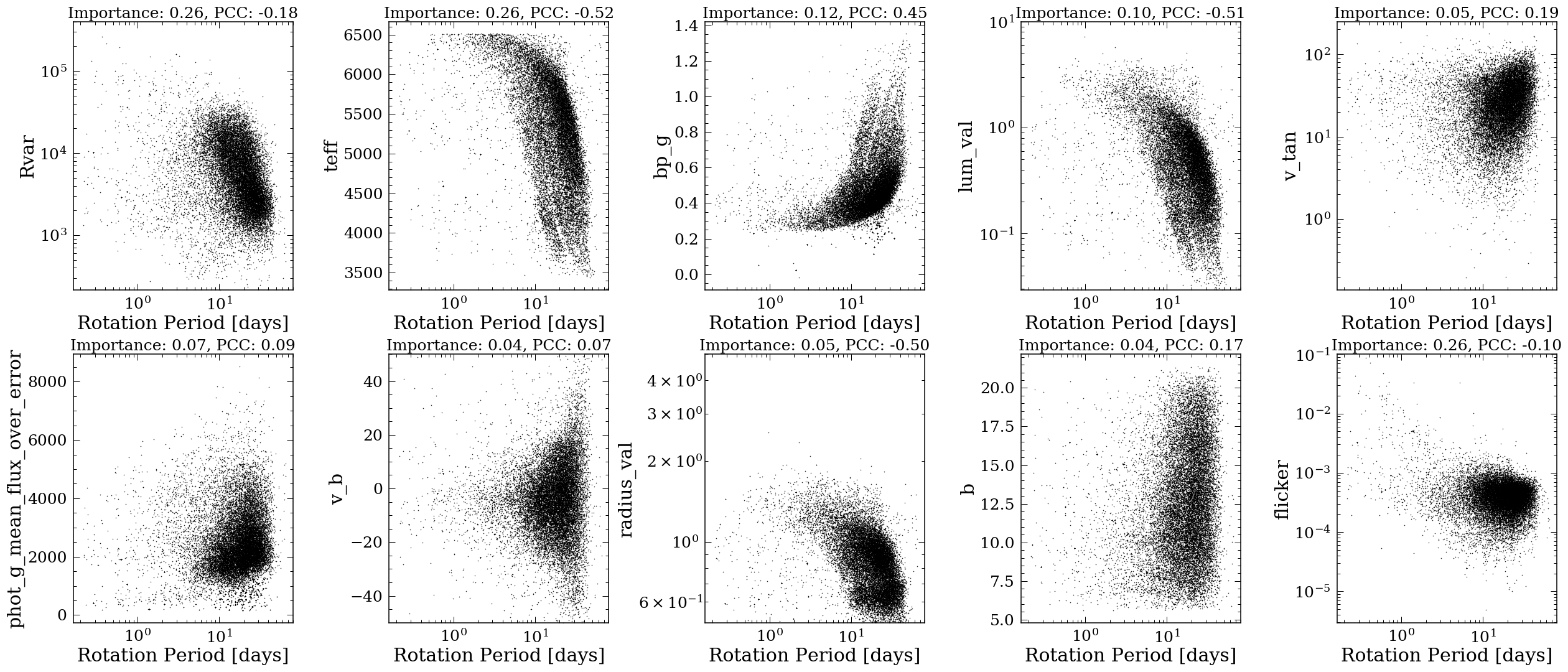}
    \caption{Relationships between features used to train the RF regressor and rotation periods for all 34,030 stars in \cite{McQuillan2014}, \cite{Santos2019} and \cite{Garcia2014}.
    It is clear that these stellar quantities are related to the rotation period of a star, but the correlations are complex and often cannot be described with simple low-order polynomials.
    The \textit{y}-axis labels are described in Table~\ref{tab:feature}.
    The title of each plot shows the {\it gini} importance (0 $\sim$ 1, where 0 is not important at all) and the Pearson correlation coefficient (PCC) for each of the features in the final training phase with the optimized hyper-parameters, which we will describe in the later sections.}
    \label{fig:features}
\end{figure*}

Fortunately, $R_{var}$ (95th percentile - 5th percentile of the normalized flux) is a good estimator for $R_{per}$ and $S_{ph}$ and its measurement does not require the information of stellar rotation.
%However, there are still some differences worth pointing out.
%There is a slight systematic offset between $R_{var}$ and $R_{per}$ from the equality line (left figure) at larger values.
$R_{per}$ is calculated by computing the 5th-95th percentile range of flux of each full stellar revolution, and then taking the average of these quantities.
On the other hand, $R_{var}$ is the 5th-95th percentile flux range of the entire light curve. 
$R_{var}$ and $R_{per}$ are therefore most similar when the stellar rotation period is long, because fewer full revolutions take place. 
Stars with long periods usually have smaller variability amplitudes, and therefore smaller $R_{var}$ and $R_{per}$ values (e.g. see Figure~\ref{fig:features}).
This is why the two quantities are similar at small values.
$R_{var}$ is more sensitive to long-term light curve systematics than $R_{per}$, and this is particularly true for rapid rotators where $R_{per}$ is calculated over many short time intervals and averaged.
This is also why $R_{var}$ is slightly larger than $R_{per}$ at large values (i.e. for rapid rotators): long-term light curve systematics slightly increase the variance in the light curve which inflates $R_{var}$ relative to $R_{per}$.
%The constant offset between $S_{ph}$ and $R_{var}$ (right) is due to the fact that $S_{ph}$ is calculated as the standard deviation instead of the 5th-95th percentile range of the light curves.
This could potentially mean we will be able to predict long rotation periods better than short rotation periods since $R_{var}$ is a better proxy to $R_{per}$ and $S_{ph}$ for slow rotators.

%\begin{figure*}[]
%    \centering
%    \includegraphics[width=\textwidth]{figures/Rvar_RperSph.png}
%    \caption{Comparisons between $R_{var}$ and $R_{per}$/$S_{ph}$.
%    Although the relations cannot be described by the identity function alone, there exists a strong correlation between $R_{var}$ and $R_{per}$/$S_{ph}$, and thus, using $R_{var}$ in replacement of $R_{per}$/$S_{ph}$ is reasonable.
%    For low $R_{var}$ and $R_{per}$/$S_{ph}$ values, these two calculations are almost identical.
%    Even for higher values ($>$ $10^4$), $R_{var}$ is still a good estimate of $R_{per}$/$S_{ph}$.
%    $\sim$ 33,000 stars and $\sim$ 15,000 stars are shown in the left and right plot respectively.}
%    \label{fig:RvarRph}
%\end{figure*}

\textit{Features Used ---}
The features used to train/test the models are (i) 3 measurements directly from the light curves, (ii) all the {\it Gaia} columns (including error columns), in which 9 were later found useful in predicting stellar rotations, and (iii) 2 kinematic statistics derived from {\it Gaia} parameters.

To obtain {\it Gaia} parameters of our sample of 133,030 {\it Kepler} stars, we used the publicly available \textit{Kepler}--\textit{Gaia} DR2 crossmatched catalog.\footnote{Available at gaia-kepler.fun}
The majority of stellar features used for rotation period prediction were obtained from the {\it Gaia} DR2 catalog, and the distance measurements were obtained from \cite{Bailer2018}.

In addition to the features from Gaia, we also calculated 3 variables directly from the light curves and 2 additional kinematic features.
These features have been shown to be related to the rotation period of a star (details described later in this section).

The features measured from the light curves are: 
(i) $R_{var}$, the range of variability in the light curve, which was calculated as the difference between the flux values at the 95th percentile and the 5th percentile,
(ii) flicker, brightness variation on timescales of 8-hours and less, calculated with \texttt{FLICKER}, our new open-source software we developed to calculate flicker using the method described in \cite{Bastien2013} (detailed description in section \ref{sec:software}), and
(iii) Lomb--Scargle periodogram maximum peak height.

Additional kinematic features we calculated are v\_tan, the velocity tangential to the celestial sphere, and v\_b, the velocity in the direction of galactic latitude from {\it Gaia} R.A. and decl. coordinates, proper motion, and parallax.

\textit{Selecting Training Features ---} \label{sec:trainingf}
To start with, our full set of features consisted of every column in the {\it Gaia} DR2 catalog, plus the three light curve statistics and the two velocities described above, making a total of 148 features.
However, we did not expect that every feature in the {\it Gaia} DR2 data set would be useful.
For example it seems unlikely the right ascension and declination would be strongly related to stellar rotation period.
Thus, we performed feature selection (selecting the important features to speed up the training process and potentially increase model performance) for both the classifier and the regressor using the method described in the next paragraph to isolate features that provide significant information about rotation periods.

We selected these features by first training the RF models on all columns from Gaia, the kinematic features and the light curve statistics calculated from 4-year stitched {\it Kepler} light curves.
We then calculating the ``{\it gini}'' feature importance \citep{breiman1984classification} using {\tt scikit-learn} \citep{scikit-learn}.
This importance was determined by calculating the mean decrease in impurity (MDI), which indicates whether a single feature alone can predict the outcome.
For example, if one can predict the rotation period of a star just by the effective temperature, then the node, where the data split (refer to section \ref{sec:intro} for how RF works), is considered pure since the model will only split the data into different subsets based on the effective temperature.
On the other hand, if the rotation period is also related to other parameters (the data is split based on more than the effective temperature), then there is an impurity in the node.
The {\it gini} importance is normalized over all features and ranges from 0 to 1.
A {\it gini} importance of 1 for a feature means the prediction of rotation period can be determined solely by this one feature.
Typically, feature values with wildly different ranges need to be normalized to a common scale in order to ensure the feature importance does not appear to be higher/lower than they should because of their range. 
However, the RF algorithm does not require feature normalization since it splits the data based on the feature values and the splitting is independent of the feature range.
Calculating this importance is a good way to eliminate irrelevant features --- features that do not contribute significantly to the prediction of stellar rotation ({\it gini} importance of 0).
We sorted the features by decreasing {\it gini} importance and performed cross-validation tests using RF regression with an increasing number of features, and selected the smallest number of features that led to a converged accuracy (for classifier) or $\chi^2$ value (for regressor).
The accuracy/$\chi^2$ value converges when the change is smaller than 5\%.
The accuracy is a way to estimate the performance of a classifier and the $\chi^2$ value is a way to estimate the deviation between the target rotation period and the predicted rotation period for a regressor.
Cross-validation tests are often used to maximize the performance of the model.
We trained the RF model on the training set, and by maximizing the model performance on the cross-validation set, we will be able to optimize the model.
To perform the cross-validation tests, we randomly excluded 20\% of the data in the training phase and predicted the rotation periods for these stars using the trained model. 
For each set of features, we performed the cross-validation test 10 times and took the median of the average $\chi^2$ values.
Figure~\ref{fig:features} shows the relationships between these features and rotation periods for the 34,030 stars in \cite{McQuillan2014}, \cite{Santos2019} and \cite{Garcia2014} as well as their {\it gini} importance.

Looking at the relationships in Figure~\ref{fig:features}, the {\it gini} importance, and the Pearson correlation coefficient (PCC, a statistical value to measure the linear correlation between two variables), there exists strong correlations between rotation period and $R_{var}$, effective temperature, \textit{Gaia} color ($G_{BP} - G$, also called bp\_g), luminosity, and radius.
There also exist weak correlations between rotation period and the other features plotted.

$R_{var}$ is known to be strongly correlated with rotation period \citep{McQuillan2014,Santos2019,Pizzolato2003,Hartman2011,Walkowicz2013}. 
It is also proven that the rotation period is a strong function of effective temperature and age \citep[This is the principle behind gyrochronology, e.g.][]{Skumanich1972, Kawaler1988, Barnes2003, Barnes2007} and age is weakly correlated with multiple stellar parameters such as luminosity, surface gravity, and kinematics.
It makes sense therefore, that $R_{var}$, effective temperature and color would have the strongest correlation with rotation period and the other stellar parameters would have weaker relations with rotation period. 

There exist both strong and weak correlations between rotation period and a number of other stellar parameters.
These relationships are difficult to reproduce using physical, or simple empirical models.
However, a machine learning algorithm like RF is effective at predicting properties from a large number of weakly correlated features, and this is why it is so well suited to predicting rotation periods from other stellar parameters.

\begin{table*}
    \centering

    \begin{tabularx}{\textwidth}{l|X|X}
    
    \hline
    Feature name & Description & Categories   \\
    \hline
    \hline
    
    bp\_g (c,r) & Integrated BP mean magnitude - G-band mean magnitude. &  Direct {\it Gaia} observations (gaia-kepler.fun). \\
    phot\_g\_mean\_flux\_over\_error (c,r) &  Mean flux in the G-band divided by its error. \\
    parallax (c)  & Parallax.  \\
    \hline
    $\mathrm{T_{eff}}$ (r) & Estimate of effective temperature from Apsis-Priam \citep{Andrae2018}. &  Stellar properties derived from {\it Gaia} observations (gaia-kepler.fun). \\
    lum\_val (r) & Estimate of luminosity from Apsis-FLAME \citep{Andrae2018}. \\
    radius\_val (r) & Estimate of radius from Apsis-FLAME \citep{Andrae2018}. \\
    r\_lo/r\_hi (c) & 68\% confidence interval on distances from \cite{Bailer2018}. \\

    \hline
    v\_tan (r) & Velocity tangential to the celestial sphere ($\sqrt{v_\mathrm{ra}^2+v_\mathrm{dec}^2}$). & Kinematic derived from {\it Gaia} proper motion, ra, dec and parallax using astropy.\\
    
    v\_b (r) & Velocity in the direction of galactic latitude.\\
    
    b (r) & Galactic latitude of the object at reference epoch \citep{Butkevich2014}.  \\
    \hline
    LG\_peaks (c) & Maximum peak height from Lomb--Scargle Periodogram. & Light curve statistics.\\
    $R_{var}$ (c,r) & Photometric variability of the light curve (95th percentile - 5th percentile of the normalized flux). \\
    flicker (r) & Brightness variation on timescales of 8 hours and less calculated with $FLICKER$ software. \\
    
    \hline
    \end{tabularx}
    \caption{Final training features used in this project sorted into four categories. Other than the radius value itself, we also included the 68\% confidence interval of distances (r\_hi/r\_low) from \cite{Bailer2018} as training features}.
    ``c'' and ``r'' represent whether the feature was used in training the classifier and regressor, respectively.
    \label{tab:feature}
\end{table*}

\subsection{Random Forest Classification and Regression} \label{sec:RF}
The RF algorithm merges multiple decision trees to get a more accurate and stable prediction.
This algorithm is also known to reduce over-fitting, which is a common problem in single decision trees.
RF can be used in both classification and regression.
It also requires less computational time compared to deep learning and is able to handle outliers.
However, RFs are not capable of extrapolating data so we could only in theory predict rotation periods up to $\sim$ 150 days, which was determined by the upper limit for rotation periods in our training set.
We used the Python \texttt{scikit-learn} package to train our RF classifier and regressor.
The hyper-parameters were set to default for the classifier for simplicity and we explored the hyper-parameters used in our regressor later on in this section.

{\it Random Forest Classification ---}
RF models are not good at extrapolating data.
This means we are only able to predict rotation periods in the same parameter space as the training set and this is the main reason we need a classifier --- to determine whether a star lies in the same parameter space as the stars in the catalog from \cite{McQuillan2014}.
Another motivation for a classifier is that not all stars with rotation periods show detectable signals in their light curve.
For example, a star could be inactive and therefore not have detectable spot modulations.
It could have starspots distributed homogeneously on the surface that cancel out any variations in its light curve.
We could also be viewing the star pole-on and, therefore, not be able to detect any azimuthal variations. 
Both of these factors require us to train a classifier to first determine if it is possible to predict a reliable rotation period.
The labels were created using stars in the \citet{McQuillan2014} catalog, where the 34,030 stars that have rotation periods were labeled ``measurable'' and the remaining 99,000 stars were labeled ``unmeasurable''.
Since the method of \citet{McQuillan2014} was conservative, our classifier trained on this data set was also on the conservative side, i.e. it is possible that the periods of some stars with periodic brightness variations in their light curves were classified as ``unmeasurable'' with our classifier.
So this classifier is not a perfect tool to determine whether a star has a detectable period but rather a way to classify whether the star would appear in the \citet{McQuillan2014} catalog. 

Features used to train the classifier: LG\_peaks, $R_{var}$, bp\_g, phot\_g\_mean\_flux\_over\_error, r\_lo, r\_hi, parallax (refer to Table~\ref{tab:feature} for detail description for each variable)

{\it Random Forest Regression ---}.
To predict rotation periods, the regressor was used if a star's period was labeled as ``measurable'' by the classifier.
Here we used a RF regressor since a star's rotation period is correlated with its other stellar properties, and RF regression is useful for predicting continuous values from various features.
A RF regressor trains multiple independent decision trees on a different subset of the data where each tree could give a slightly different period prediction.
The model then takes the average of all the predictions from all the trees and their uncertainties to determine the final predicted rotation period.

Features used to train the regressor: $R_{var}$, teff, bp\_g, lum\_val, flicker, radius\_val, v\_tan, phot\_g\_mean\_flux\_over\_error, b, v\_b (refer to Table~\ref{tab:feature} for detail description for each variable)

\section{Optimizing and assessing the performance of the Random Forest models}\label{sec:Optimize}

We trained both the classifier and the regressor on 80\% of the data and used the remaining 20\% to perform cross-validation tests, which is a good way to prevent over-fitting.
The features used for each model and their permutation feature importance are shown in Figure~\ref{fig:Imp}.

\subsection{Random Forest Classifier}
\begin{figure}[h!]
    \centering
    \includegraphics[width=0.45\textwidth]{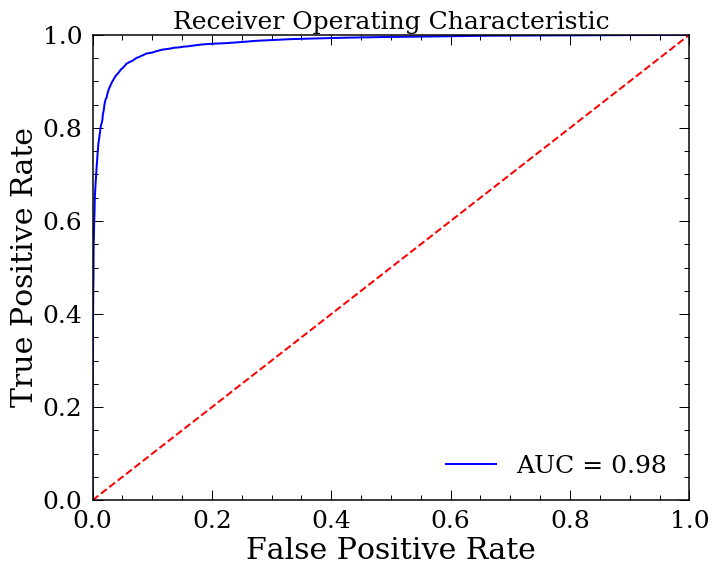}
    \caption{The Receiver operating characteristic curve for the classifier.
    The Area Under The Curve (AUC) shows how well the model can distinguish between classes.
    It is often used to estimate the accuracy or performance of the classifier.
    The model reached maximum accuracy of 98\% with a threshold of 0.4. }
    \label{fig:ROC}
\end{figure}

The outputs of the classifier were numbers from 0-1 for each star, where 0 means the period was 100\% ``unmeasurable'' and 1 means it was 100\% ``measurable''.
One can simply say if the predicted number was greater than 0.5 (which means the threshold was 0.5) then the period was ``measurable''.
However, the best way to determine the threshold is to maximize the area underneath a Receiver Operating Characteristic curve (ROC) as shown in Figure~\ref{fig:ROC}.
A ROC curve shows the predicted False Positive Rate (FPR) against the True Positive Rate (TPR) for various thresholds.
The FPR is the total number of False Positive cases (e.g. number of stars where their rotation periods can be measured and are predicted ``unmeasureable'') divided by the total number of negative cases (e.g. number of stars where their rotation periods are predicted ``unmeasureable'').
The TPR is the total number of True Positive cases (e.g. number of stars where their rotation periods can be measured and are predicted ``measureable'') divided by the total number of positive cases (e.g. number of stars where their rotation periods are predicted ``measureable'').
These statistics are useful especially in cases where the training data set is overflowed by one label (positive or negative).
For example, if 98\% of the stars in the training set has ``measurable'' rotation period, then a incorrect model that predicts every star has a ``measureable'' rotation period will reach an accuracy of 98\%.
However, this model is clearly wrong, and by calculating the TPR and FPR, one can get a better understanding of the true accuracy of the model. 
A perfect model would have a false positive rate of 0 and a true positive rate of 1 and the curve would go straight up the TPR axis until it reaches 1 then go horizontal on the FPR axis.
Thus, the closer the ROC curve approaches (0,1), the more accurate the classifier is. 
We determined the accuracy and threshold by finding the point along the curve where TPR-FPR was maximized.
This yielded a 98\% accuracy with a threshold of 0.4. 

\subsection{Random Forest Regressor}
\textit{Hyper-parameter optimization}.
To achieve the best performance of the model, we optimized the hyper-parameters (parameters describing the model) using a grid search method.
The hyper-parameters we considered and their optimal values are shown in Table ~\ref{tab:hyper}.
For each set of hyper-parameters, we performed a Monte-Carlo cross-validation test 10 times with 20\% of the data, chosen randomly each time, left out during the training process.
For each of these tests, we calculated the average $\chi^2$ = $\frac{\Sigma_i^N(y_i-y_{predict})^2/\sigma^2_{y_i}}{N}$ and the relative median absolute deviation (rMAD) = $median(\frac{|y_i-y_{i_{predict}}|}{y_i})$, where $(i = 1, 2, 3, ..., N)$, $y_i$ is the expected rotation period value, $y_{predict}$ is the predicted rotation period value and $N$ is the number of data points.
The overall average $\chi^2$ and rMAD of the model for each set of hyper-parameters were then represented by the median values of these 10 tests.

Two sets of optimal hyper-parameters were obtained by minimizing the average $\chi^2$ or minimizing the rMAD.
Minimizing $\chi^2$ reduced the spread of the data (variance) and minimizing the rMAD reduced the systematic bias in the data (bias).
In order to get a more {\it precise} result, we selected the model that minimized the average $\chi^2$.

\begin{table*}
    \centering

    \begin{tabularx}{\textwidth}{l|X|X|X|X}
    
    \hline
    hyper-parameter name & Description & Grid-search range & Value that minimizes average $\chi^2$ & Value that minimizes average rMAD   \\
    \hline
    $\mathit{n\_estimator}$ & \# of decision trees used in the RF model & 1-100 & 20 & 1 \\
    
    \hline
    $\mathit{max\_depth}$ & Maximum depth of the tree & 1-150 & 50 & 100 \\
    
    \hline
    $\mathit{max\_features}$ & \# of features to consider when looking for the best split & 1-10 & 6 & 3 \\
    
    \hline
    \end{tabularx}
    
    \caption{Optimal hyper-parameters in the RF regressor model that minimized the average $\chi^2$ or rMAD.
    By minimizing the average $\chi^2$, we had low variance but high bias in the model, and by minimizing the MAD, high variance but low bias was achieved. }
    \label{tab:hyper}
\end{table*}

\begin{figure*}
    \centering
    \includegraphics[width=\textwidth]{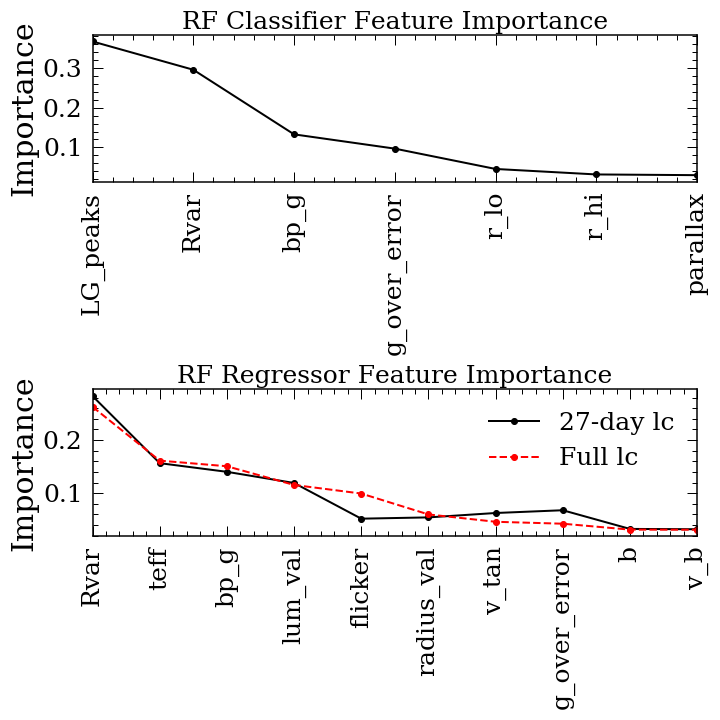}
    \caption{Permutation feature importance on the {\it Kepler} cross-validation set ($\sim$ 7,000 stars), where \textit{g\_over\_error} is the G-band mean flux divided by its error.
    The two lines in the regressor represented the feature importance for training on 27-day light curves (solid black line) and that on full 4-year {\it Kepler} light curves (dashed red line).
    $R_{var}$ and flicker are measurements for {\it Kepler} light curves with their corresponding time-scale (measured from 27-day light curves for the solid black line and from 4-year light curves for the dashed red line).}
    \label{fig:Imp}
\end{figure*}

\textit{Permutation feature importance ---}
We calculated the permutation feature importance to study how each feature impacts the prediction results using the optimized model.
By calculating the permutation feature importance, we are able to interpret the model and potentially gain insight on how the stellar properties are related to the rotation period. 
The permutation feature importance can be calculated by randomly shuffling values within a single feature and observing how the model performance changes.
This is effectively removing each feature from the model and preventing it from being informative and measure how good the model can still predict the data.
This importance is a more accurate measurement of how much of a role each feature plays in determining the outcome, compared to the {\it gini} importance. 
We used the $R^2$ (coefficient of determination) regression score to measure the model performance, $R^2 = \Sigma_i(y_{predict}-\overline{y})^2/\Sigma_i(y_i-\overline{y})^2$, where $y_{predict}$ is the predicted rotation period and $\overline{y}$ is the average rotation period. 
It provides a measure of how close the data are to the fitted regression function.
The $R^2$ score is commonly between 0 and 1, and the higher the score, the better the fit is. 
To obtain the importance for each feature, we calculated the $R^2$ score on the training set and re-shuffled the values within one feature and kept the rest of the training data set unchanged.
We then passed the new training data set to the model again to calculate a new score based on this modified training set.
The feature importance is the difference between these two scores, normalized to sum to one across features.
%The values were then normalized so the sum was 1.
Figure~\ref{fig:Imp} shows the permutation feature importance for both the RF classifier (using 4-year {\it Kepler} light curves) and the RF regressor (separately calculated for 4-year {\it Kepler} light curves and 27-day {\it Kepler} light curve segments).

The power of the highest peak in the Lomb--Scargle periodogram of each star's light curve (LG\_peaks) was the most important feature for the classifier.
Since the classifier was trained on targets from \cite{McQuillan2014}, the RF classifier learned the algorithm they used to determine whether the light curve signal was periodic.
\cite{McQuillan2014} determined whether a rotation period was reliable (or whether if a star has rotation period signal that can be detected) by setting a threshold for the maximum peak height from the ACF, which is similar to the maximum peak height from the Lomb--Scargle periodogram.
As a result, it makes sense that LG\_peaks is the most important feature in determining whether a star can be included in the catalog from \cite{McQuillan2014}.

The confidence interval of the distances also determines whether a star's period is measurable or not.
One potential reason is that a larger distance error (or any error from luminosity, temperature etc.) is also associated with a larger error in the observables (photometry and parallax, etc.), suggesting fainter or/and more distant star whose period would normally be harder to determine. 
Since errors from stellar properties are correlated \citep{Andrae2018}, the RF classifier would only use one of these errors as an important feature (similar to determining the rotation period, the RF regressor treated the effective temperature as one of most important features but not the color, though they are very similar). 
 
Other features, such as $R_{var}$, $bp\_g$, $g\_over\_error$ and distances, not only determine whether we can recover the rotation period or not but also contain information about the rotation period itself.
Since a shorter rotation period is easier to recover, it is not surprising that these attributes appear to be important in both classification and regression models. 

The importance of the regressor was more evenly distributed over multiple features.
This implies that the rotation period is closely related to multiple stellar properties and precise rotation periods can only be predicted using multiple features.
$T_\mathrm{eff}$ and $R_\mathrm{var}$ are known to be strongly correlated with rotation periods \citep[\eg][]{Santos2019}, and the kinematics of a star, as mentioned in section \ref{sec:intro}, could also be used to constrain its age, and therefore, its period.

The importance trend for the model trained on 27-day {\it Kepler} light curves closely follows the model trained on full 4-year {\it Kepler} light curves.
However, the flicker feature is more important for 4-year light curves compared to that of 27-day light curves.
This suggests the flicker value encodes more information from the rotation period as we average over longer time-spans or that the flicker measurement becomes more precise, and therefore more discerning of logg, as more quarters are incorporated.

\section{Results}\label{sec:results}
In this section, we present the performance of our optimized model (M\_$\chi^2$) on {\it Kepler} data with 4-year light curves, on simulated {\it TESS} data, calculated by splitting full {\it Kepler} light curves into 27-day sections, and on real {\it TESS} data.

\subsection{Performance on {\it Kepler} Data}
\begin{figure*}
    \centering
    \includegraphics[width=\textwidth]{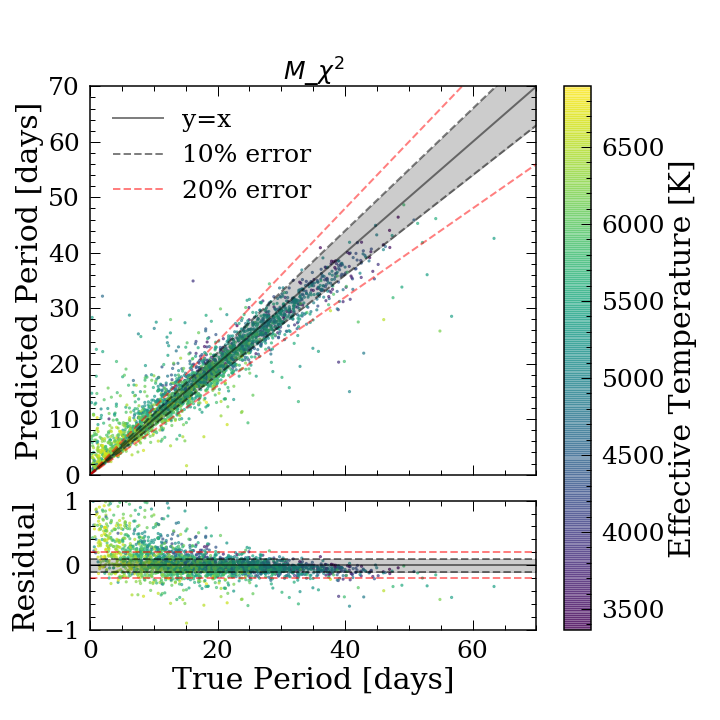}
    \caption{Periods predicted by RF regressor versus the rotation periods measured in \cite{McQuillan2014} for $\sim$ 3,000 stars colored by their effective temperature, where hotter stars tend to rotate faster, as expected. 
    The grey area occupies the 10\% error space.
    The top panel shows the predicted rotation period vs true rotation period from \cite{McQuillan2014} and the bottom panel shows the residual.
    We are able to predict rotation periods with an uncertainty of 13\% and long rotation periods ($>$ 30 days) with an uncertainty of 9\%.}
    \label{fig:KeplerPredict}
\end{figure*}

Figure~\ref{fig:KeplerPredict} shows the testing result on full 4-year {\it Kepler} light curves with points colored by their effective temperature. 
In general, cooler stars spin more slowly because they have deeper convection zones which means they have stronger magnetic fields and therefore spin down faster due to magnetic breaking compared to hotter stars.
We picked the model with the lowest $\chi^2$, which also minimized the scatter (variance).
However, low variance models normally have high systematic bias.
It is clear from the residual shown in the bottom panel that we systematically over-predicted the short rotation periods and under-predicted the long rotation periods.
We estimated the uncertainty by calculating 1.5*MAD from the residuals and we can recover the rotation periods with an uncertainty of 13\% and long rotation periods ($>$ 30 days) with an uncertainty of 9\%.

\subsection{Performance on 27-day {\it Kepler} light curves}
Testing our model on {\it Kepler} 4-year light curves gave us promising results.
However, our main goal for this model is to predict rotation periods from 27-day {\it TESS} light curves.
To do this, we split each 4-year light curve from the {\it Kepler} training set into multiple 27-day segments and calculated $R_{var}$ and flicker for these short-duration light curves.
Other features from {\it Gaia} remained the same for each target.
Breaking up the light curves from the {\it Kepler} training set, and treating each 27-day light curve as a separate star, effectively expanded our number of training targets to over 1.8 million ($\sim$ 34,000 4-year light curves from Kepler, with each of these light curves splitting into $\sim$ 54 27-day light curves). 

\begin{figure}[h!]
    \centering
    \includegraphics[width=0.45\textwidth]{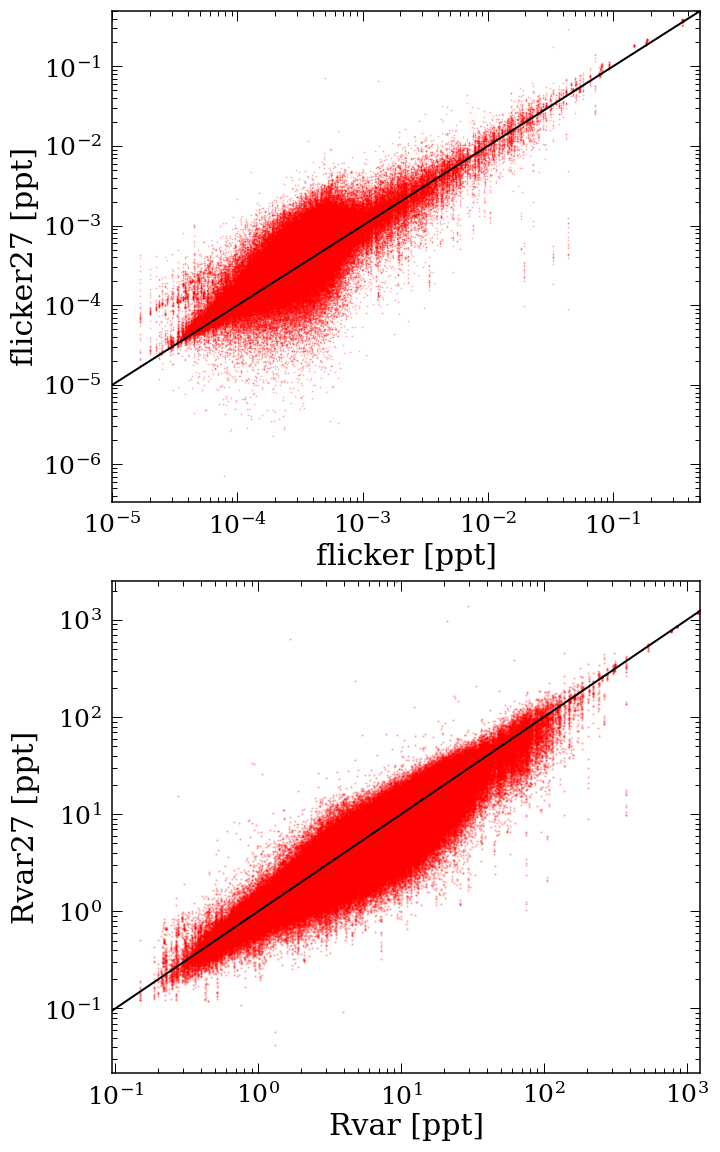}
    \caption{Flicker and $R_{var}$ values for 4-year light curves compared to those of 27-day lightcuves for $\sim$ 100,000 {\it Kepler} stars ($\sim$ 5 million 27-day light curve statistics from splitting up 4-year light curves). The solid lines are the identity functions. The relative MAD between 27-day light curves and 4-year light curves are 30\% and 35\% for flicker and $R_{var}$ respectively.}
    \label{fig:Rvarflicker}
\end{figure}

{\it Comparing 4-year and 27-day light curves ---}
Figure~\ref{fig:Rvarflicker} shows a comparison between $R_{var}$ and flicker values from 4-year light curves with those of 27-day light curves.
We quantified the differences between these two statistics for the 4-year light curves and 27-day light curves by calculating 1.5*MAD (a measure of the standard deviation that is robust to outliers) of the residuals.
This yielded a standard deviation of 30\% and 35\% for flicker and $R_{var}$, respectively. 
The scatter in these two light curve statistics constrains how well we can predict rotation periods and is discussed in the later paragraphs. 

\begin{figure*}
    \centering
    \includegraphics[width=\textwidth]{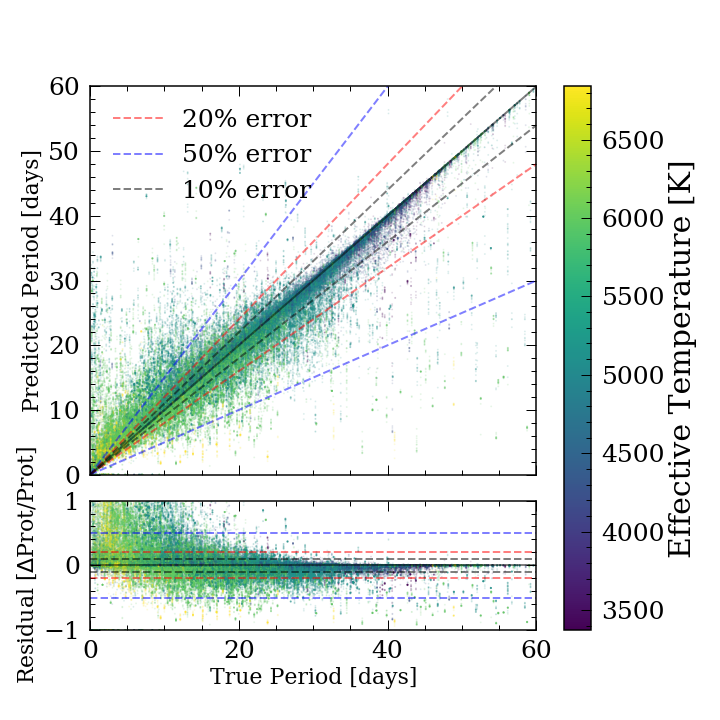}
    \caption{Results for 370,208 27-day {\it Kepler} light curve segments, colored by {\it Gaia} effective temperature.
    The top panel shows the comparison between predicted rotation periods and true rotation periods.
    The bottom panel shows the residual, 3\% of the data were cut out that has residuals greater than 1.
    There is a clear temperature gradient from fast rotators to slow rotators, where hotter stars tend to rotate faster, as expected.
    The period predictions have an average uncertainty of 9\% for all the stars and 5\% for slow rotators ($>$ 30 days).}
    \label{fig:KeplerCV}
\end{figure*}

After excluding the 195 stars observed by both {\it Kepler} and {\it TESS}, which we later tested our model on, we trained the model on $\sim$80\% of these 1.8 million 27-day light curves and tried to recover the remaining 20\%.
Figure ~\ref{fig:KeplerCV} shows the results for $\sim$20\% of the targets ($\sim$ 300,000) in the \citet{McQuillan2014} catalog.
We did not optimize the hyper-parameters again since the the both training sets are from Kepler and we assumed the light curve statistics we calculated will be similar so the optimized hyper-parameters will also be similar. 
The general trend follows that shown in Figure~\ref{fig:KeplerPredict}.
But with more training data (since we broke the {\it Kepler} 4-year light curves into multiple 27-day light curves), most of the predictions have uncertainties on the order of 9\%, and we are able to predict long rotation periods ($>$ 30 days) with an uncertainty of 5\%.
This is important because despite the measurements for $flicker$ and $R_{var}$ from 27-day light curves being worse, we were able to get a more precise result by increasing the number of training data by splitting the full 4-year light curves. 
A fit could potentially be used to correct for the bias, however, this bias is subject to change.
For example, the difference in the noise properties between {\it TESS} and {\it Kepler} could affect the systematic bias.
More discussion is included in section \ref{sec:disc}.

One additional feature worth pointing out is the vertical streaks in Figure~\ref{fig:KeplerCV}.
This is most likely due to the variation in $R_{var}$ and flicker (Figure~\ref{fig:Rvarflicker}). 
After splitting the 4-year light curve of each star into multiple 27-day light curves, there existed multiple training data that had the same values for every feature except $R_{var}$ and flicker (since we recalculated these two values for every 27-day light curve).
This would cause the model to have multiple different predictions for a same star even though this star only has one rotation period measured with traditional methods.
    
One concern is that we were not able to recover rotation periods for fast rotators with high precision when compared with the use of traditional methods.
One potential reason could be that some very fast rotators are synchronized binaries.
Synchronized binaries are binary stars whose tidal interactions have synchronized their rotation periods with their orbital periods, i.e. they are tidally locked with each other.
There is mounting evidence to show that a large fraction of cool stars which rotate faster than 7-10 days are, in fact, synchronized binaries \citep[e.g.]{Angus2020,simonian2019}.
The rotation periods of stars in synchronized binary systems have been influenced by tides, and will not be the same as (and will probably be shorter than) the rotation period expected for each star based on their temperatures, surface gravities, and ages.

Our main goal with this RF model was to predict long rotation periods with short {\it TESS} light curves which is difficult to do using traditional methods.
So not being able to predict short rotation periods accurately is not a major concern for our algorithm since we could combine both methods to measure rotation periods of all ranges.
Furthermore, we are predicting rotation periods instead of measuring.
This means even though our results are not as accurate as periods measured with traditional methods, we can still predict rotation periods when traditional methods fail to measure.

\subsection{Performance on real {\it TESS} data}\label{sec:TESS}
We downloaded the 195 {\it TESS} 2-min cadence PDCSAP light curves from MAST and calculated the Lomb--Scargle maximum peak height, flicker, and $R_{var}$ from the {\it TESS} 27-day light curves.
The rest of the features were acquired from Gaia.
We first passed these stars through the trained classifier and all 195 rotation periods were identified as ``measurable''.
These targets were then fed to the trained regressor (trained on 27-day light curves) in order to predict their rotation periods.

Ideally, we would train the model on {\it TESS} targets since the variables calculated from the light curves (Flicker/$R_{var}$) are expected to differ between {\it TESS} and {\it Kepler} due to their different bandpasses.
Detailed discussions of the difficulties of applying a model trained on {\it Kepler} to {\it TESS} are included in section \ref{sec:disc}.
However, we do not yet have a large enough training set for {\it TESS} that includes enough rotation periods.
Because of that, the result here is a preliminary test of how well the model, trained on Kepler, can predict rotation periods from {\it TESS} short-duration light curves.  

\begin{figure*}
    \centering
    \includegraphics[width=\textwidth]{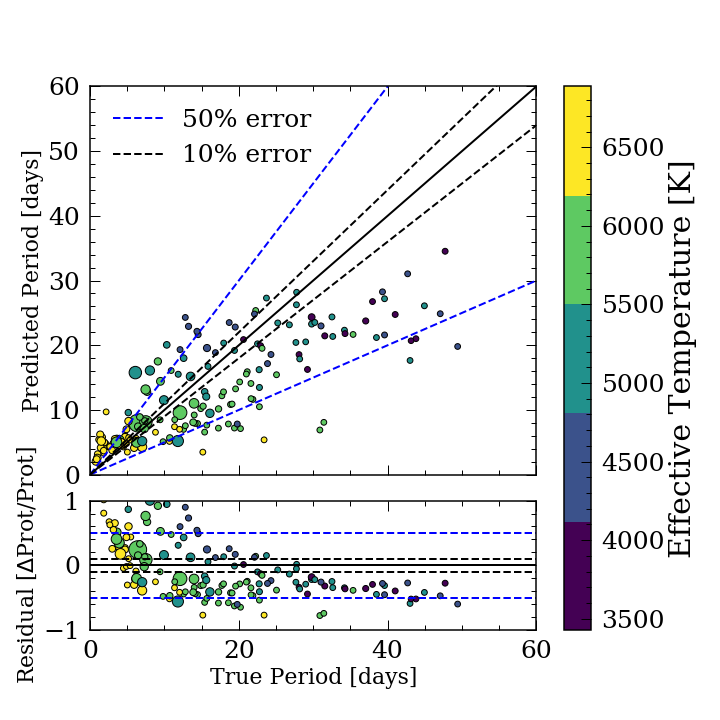}
    \caption{Testing result on 195 {\it TESS} targets in the {\it Kepler} field using the M\_$\chi^2$ model trained on 27-day {\it Kepler} light curve segments, colored based on the effective temperature. The marker size indicates the signal-to-noise ratio (S/N). The S/N is calculated by dividing $R_{var}$ by the noise floor level calculated in Figure~\ref{fig:TKlc}. The uncertainty is 55\% for all predictions as well as for slow rotators.}
    \label{fig:TESSKepler}
\end{figure*}

The major difference between the results for simulated and real {\it TESS} light curves (Figure~\ref{fig:KeplerCV}  and \ref{fig:TESSKepler}, respectively), is that the model, tested on real {\it TESS} data, suffers from higher bias for slow rotators.
This may be due to additional white noise scatter in {\it TESS} light curves, which limits measurements of $R_{var}$ and flicker in real {\it TESS} light curves.
The signal-to-noise ratio for $R_{var}$ (indicated by the size of the marker, the larger the marker, the higher the S/N) indicated that the summery statistics calculated from the {\it TESS} light curves are not reliable and are possibly dominated by the noise (further discussion in section \ref{sec:disc}).
As a result, the predictions are most likely dominated by the temperature of the star, which is supported by the clear color gradient. 
However, this preliminary test shows promising results in using RF to predict long rotation periods from short-duration light curves from \textit{TESS}.
%%As shown in Figure~\ref{fig:TESSKepler}, the results for real {\it TESS} stars have several similar major characteristics to the results for 27-day {\it Kepler} light curve segments in Figure~\ref{fig:KeplerCV}:
%%\begin{itemize}
%%    \item Stars that have medium to short rotation periods ($<$ 30 days) shows similar behaviors as the result for the 27-day {\it Kepler} light curve segments.
%%    \item The model suffers from high bias, especially for slow rotators.
%%    One difference worth pointing out is that the systematic bias in the model on simulated {\it TESS} data and real {\it TESS} data deviates drastically.
%%    This is expected since the light curve properties for {\it TESS} and {\it Kepler} are very different (further discussion in section \ref{sec:disc}).
%%    \item The preliminary test shows promising results in using RF to predict long rotation periods from short-duration light curves from TESS.
%%\end{itemize}

\section{Discussion \& Future Work}\label{sec:disc}
In performing this analysis, we revealed a few limitations and unforeseen possibilities for our random forest classification and period prediction.
We outline the most important of these below.

\textit{Better long rotation period predictions for {\it Kepler} stars ---}
It is clear from the uncertainty analysis in Figure~\ref{fig:KeplerPredict} and Figure~\ref{fig:KeplerCV} that we are able to predict long rotation periods with a higher precision ($\sim$ 4\% better) than short rotation periods using the RF regressor.
There are a few reasons why this might be the case.

\begin{itemize}
  \item {\it Inhomogeneous data ---}
  We added stars with long rotation periods from \cite{Santos2019} and \cite{Garcia2014} and they did not use the same methods to determine the rotation periods as \cite{McQuillan2014}.
  \cite{Santos2019} and \cite{Garcia2014} used the combination of wavelet analysis and the ACF, whereas \cite{McQuillan2014} only used the ACF.
  Because of the differences in their methods, the rotation period measurements from \cite{Santos2019} and \cite{Garcia2014} could be slightly different than those from \cite{McQuillan2014}.
  This could cause the data splitting in the RF regressor to be biased, causing it to find a slightly different relation between features and long rotation periods, and ending up with better predictions for slow rotators. 

  \item {\it Physics of the slow rotators ---} 
  Slow rotators might have a more straightforward relationship between their stellar properties and their rotation periods.
  In Figure~\ref{fig:features}, there seems to be less scattering in $R_{var}$ versus rotation period, and $R_{var}$ is the most important feature in predicting rotation periods. 
  In addition, rotation periods for fast rotators might still be affected by initial conditions from when the stars were born. As stars contract onto the main-sequence, they gradually spin down.
  As a result, some of the fast rotators might still contain information of their birth angular momentum so their stellar properties are not closely related to their rotation periods.
\end{itemize}

{\it Information in the light curve ---}
The fact that we are able to predict long rotation periods ($>$ 27 days) by training on 27-day light curves, plus {\it Gaia} photometry, seems counter to intuition. 

However, this is demonstrative of the utility of automated methodologies like RF regressors, to learn the mapping from data to label, on a data point-by-point basis.
Similarly, \cite{Blancato2020} uses a convolutional neural network to predict stellar properties, including rotation periods, directly from 27-day {\it Kepler} light curves.
They are able to recover short rotation periods better than the method presented here for $<$ 35 day periods.
This suggests that, by calculating only a couple summary statistics, we did not use all the information contained from the light curve.
However, \cite{Blancato2020} are not able to predict rotation periods $>$ 35 days as accurately as our approach.
The comparison could also support the idea that in order to accurately predict long rotation periods from short-duration light curves, we need more than just the information contained in the light curve themselves.

\textit{Limitations of predicting {\it TESS} rotation periods ---}
There are a couple of important differences between {\it Kepler} and {\it TESS} that make applying a trained model on {\it Kepler} to {\it TESS} difficult.
Here, we discuss differences in observing direction, band-pass, precision, and cadence.

\begin{itemize}
\item Observing direction: {\it TESS} points at a different area of the sky every 27 days whereas {\it Kepler} only pointed at one direction.
The kinematics used to train the model are not in the galactic coordinates system since the radial velocities are not available for most stars.
Therefore the v\_tan and v\_b relations with age are different in different directions.
Although the kinematics were not that important for determining the rotation periods for {\it Kepler} stars (see Figure~\ref{fig:Imp}), we expect they may be more important for predicting stellar rotations for stars in the {\it TESS} observing field.
As a result, we will only be able to predict rotation periods for stars in the direction of the {\it Kepler} field.

\item Band-pass differences: {\it TESS} and {\it Kepler} also have different observing bandpass and instrumental precision.
{\it TESS} is targeting low-mass stars, which are cooler and redder, whereas {\it Kepler} is targeting sun-like stars.
As a result, {\it TESS} observes in the wavelengths of $\sim$600-1100~nm, whereas {\it Kepler} observed between the wavelengths of $\sim$400- 900~nm.
Because of this, any calculations made from the light curves (e.g. $LG\_peaks$, $R_{var}$ and \textit{Flicker}) are likely to be different.
Figure~\ref{fig:TKlc} shows comparisons between $R_{var}$ and \textit{Flicker} calculated from {\it TESS} and {\it Kepler} light curves for the 195 testing stars.
Flicker calculated from {\it TESS} is always greater than that calculated from Kepler.
This could be because the surface granulation signal of a star corresponding with flicker is louder in redder band-passes.
This would mean flicker could potentially have more information about rotation periods in the {\it TESS} light curves and be a more important feature than it was in the {\it Kepler} training features.
Alternatively, this could be because {\it TESS} light curves have higher-amplitude white noise background than {\it Kepler} light curves, which is added to the flicker estimate (see point below).
One could correct these values based on {\it TESS} magnitude and obtain a better result on the {\it TESS} test set. 

\begin{figure}[h!]
    \centering
    \includegraphics[width=0.45\textwidth]{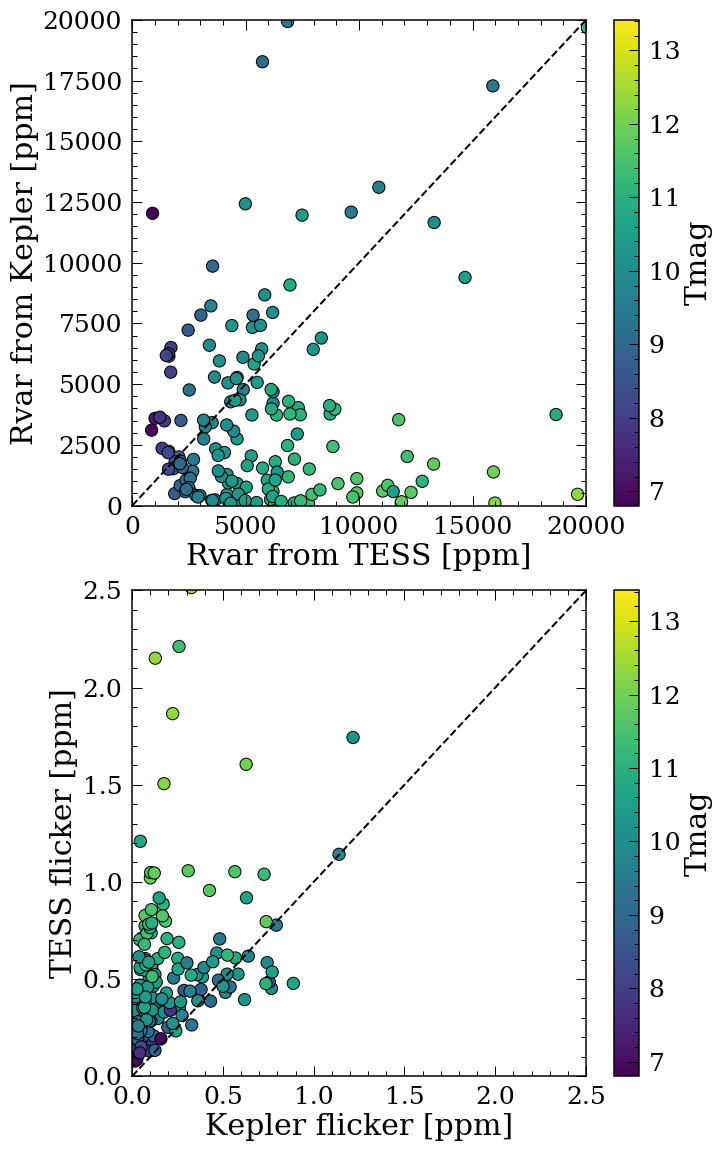}
    \caption{Comparisons between $R_{var}$ [ppm] and $Flicker$ [ppm] calculated from the {\it Kepler} light curves and {\it TESS} light curves of the 195 testing stars, colored by {\it TESS} magnitude. There is a magnitude gradient in both plots and $Flicker$ values calculated from {\it TESS} light curves are systematically higher than those of {\it Kepler} light curves.}
    \label{fig:TKlc}
\end{figure}

\item Instrumental precision: {\it TESS} has a lower instrumental precision at all magnitudes compares to that of {\it Kepler}.  
Figure~\ref{fig:TK_std} shows the systematic noise versus {\it TESS} magnitude for the {\it TESS} and {\it Kepler} light curves of the 195 testing stars.
We calculated the systematic noise for these 195 stars by measuring the standard deviation of the flux in a 3-hour window and took the median of these values. 
Although following similar trends, the systematic noise in {\it TESS} light curves is one order of magnitude higher than that of {\it Kepler} for a given {\it TESS} magnitude.

In addition to the fact that {\it TESS} has higher systematic noise in the light curves, the noise floor, especially for high {\it TESS} magnitude, is comparable to the $R_{var}$ and $flicker$ measurements (see Figure~\ref{fig:TKlc}).
This could mean that these measurements are not accurate or even worse, we could be measuring the systematic noise instead of any physical quantities.
The noise floor of {\it TESS} could also limit our ability to predict long rotation periods since stars with longer rotation periods typically exhibits lower $R_{var}$ signals.
\begin{figure}[h!]
    \centering
    \includegraphics[width=0.5\textwidth]{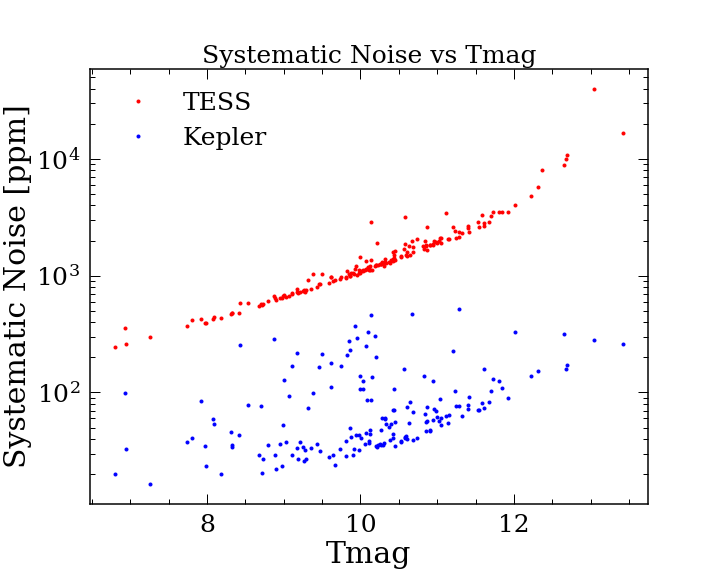}
    \caption{systematic noise (standard deviation on a 3-hour window) versus {\it TESS} magnitude for the 195 stars observed by both missions.
    At any given {\it TESS} magnitude, the systematic noise in the {\it TESS} light curve is always, on average, one magnitude higher than that of a {\it Kepler} light curve for the same star.
    This means any measurements extracted from the {\it TESS} and {\it Kepler} light curves are expected to be different.}
    \label{fig:TK_std}
\end{figure}

\item {\it Pixel size ---}
{\it TESS} has a pixel size of 21 arcseconds, which is large compared to Kepler, which has 3.98 arcsecond pixels.
This means the {\it TESS} light curves are more likely to be affected by contamination from nearby star light. 

\item Cadence: We calculated the light curve statistics ($R_{var}$ and flicker) from both the original {\it TESS} light curves (2-min cadence) and the smoothed light curve (taking the rolling median of 30 minutes to simulate {\it Kepler} cadence) and did not find significant changes.
Therefore, the differences between the cadence in {\it TESS} and {\it Kepler} would not be significant.
However, in this project, we only investigated the effects between 30-min and 2-min cadence data and extending the study to other cadence differences is interesting but beyond the scope of this project (\cite{Blancato2020} have done a more thorough study of the effect of cadence).

\end{itemize}

Despite the differences, we were still able to recover long rotation periods from real {\it TESS} light curves within 50\% uncertainty.
This means our model can potentially be applied to other surveys such as LSST and PLATO.

\textit{Potential alternative uses for this RF regressor ---}
The main goal for this model is to predict long rotation periods ($>$ 15 days) for main sequence stars from 27 day {\it TESS} light curves, however it may have other applications.
Since RF models are not particularly good at extrapolating data, any stars that have anomalous stellar parameters are most likely to be identified as outliers.
Consequently, this model could potentially be used to gain insight on the outliers within data sets.
Here, we list a couple of potential applications for this RF model:
\begin{itemize}
    \item If a star has a rotation period predicted much larger than the measured rotation period from traditional methods (LG, ACF, etc.), this star may have undergone tidal synchronization, resulting from a closely orbiting companion star.
    We could possibly create a synchronized binary detector with our regressor.
    \item We could try to infer the inclination of a star by predicting $R_{var}$ from the known rotation period.
    If a star is inclined to be almost pole-on, its photometric variability measured directly from the light curve will be smaller than that predicted for the $R_{var}$-stellar rotation relation.
    \item We could compare the rotation periods of stars with close orbiting Hot Jupiters and those without to study how these Hot Jupiters might affect the rotation period and magnetic activity of their host stars.
    
\end{itemize}

%\textit{Plotting predicted rotation periods versus {\it Gaia} temperature ---}

\textit{Future work ---} Due to the limitations of predicting {\it TESS} rotation periods with a model trained on the {\it Kepler} dataset, we will want to train our RF regressor on rotation periods measured from {\it TESS} targets across the entire observing zone using ACF.
We will then want to create a catalog of {\it TESS} rotation periods that can be used by the astronomy community. 
It would also be interesting to investigate how the sparsity in feature parameter space affects the model prediction.

\section{Conclusion}\label{sec:concl}
Rotation periods are important for studying stellar magnetic activity, improving RV measurements for exoplanet searches, and even in determine stellar ages.
Stellar rotation periods have been precisely measured using traditional methods, such as periodograms and autocorrelation functions, for {\it Kepler} targets.
However, instead of having 4-year light curves, most {\it TESS} stars will only have 27-day light curves for every one-year observing window.
This increases the difficulty of using traditional methods to recover rotation periods, especially those of M dwarfs, which often have periods greater than 27 days \citep{McQuillan2014}.

We presented a new method to predict long rotation periods from short-duration light curves using Random Forest, a machine learning algorithm.
We first trained a RF classifier on stars from \cite{McQuillan2014,Santos2019,Garcia2014}, {\it Gaia} DR2 \citep{Prusti2016, Brown2018} and distances from \cite{Bailer2018} to identify whether the rotation period of a star is ``measurable''.
A regressor, trained on the same targets, was then used if the rotation period of a star could be predicted based on the classifier.
The data set and features used to train these models were described in section \ref{sec:datamethod}.
We find that the most important features used to predict rotation periods are $R_{var}$, effective temperature, {\it Gaia} color, luminosity, and flicker.
We calculated the uncertainties by calculating the median absolute deviation of predicted rotation periods.
We were able to predict rotation periods of {\it Kepler} stars with an average uncertainty of 13\% (9\% for rotation periods $>$ 30 days) with 4-year light curves and 9\% (5\% for rotation periods $>$ 30 days) with 27-day light curves.
We found that long rotation periods were predicted more precisely than short rotation periods. 
When applying this regressor trained on {\it Kepler} data to {\it TESS} data, we were able to recover rotation periods of {\it TESS} stars in the {\it Kepler} field with an uncertainty of 55\%.
The decrease in precision was most likely due to the differences between the two missions, described in section \ref{sec:disc}.
This preliminary test on {\it TESS} stars showed promising results and we expect to be able to predict rotation periods with smaller errors if we can train the regressor on {\it TESS} targets.
The two open-source software packages, \texttt{FLICKER} and \texttt{Astraea}, developed in this project, are available on Github and are described in section \ref{sec:software}.
In the future, we hope to train the RF regressor on {\it TESS} data and create a catalog of rotation periods. 

\acknowledgments
%\section{Acknowledgement}\label{sec:ack}
R.A. acknowledges support from NASA award: 80NSSC20K1006

We want to thank Angela Santos for sharing her data.

This work made use of the gaia-kepler.fun crossmatch database created by Megan Bedell.

This paper includes data collected by the {\it Kepler} mission. Funding for the {\it Kepler} mission is provided by the NASA Science Mission directorate. 

This paper includes data collected by the {\it TESS} mission. Funding for the {\it TESS} mission is provided by the NASA Explorer Program.

This work has made use of data from the European Space Agency (ESA) mission
{\it Gaia} (\url{https://www.cosmos.esa.int/gaia}), processed by the {\it Gaia}
Data Processing and Analysis Consortium (DPAC,
\url{https://www.cosmos.esa.int/web/gaia/dpac/consortium}). Funding for the DPAC
has been provided by national institutions, in particular the institutions
participating in the {\it Gaia} Multilateral Agreement.

This research made use of Astropy,\footnote{http://www.astropy.org} a community-developed core Python package for Astronomy \citep{astropy:2013, astropy:2018}.


\vspace{5mm}
\facilities{\textit{Gaia}, \textit{Kepler}, \textit{TESS}}


%% cite Astraea bellow:
\software{Astraea (this work), Astropy \citep{astropy:2013, astropy:2018}, FLICKER (this work), Numpy \citep{oliphant2006guide}, Scikit-learn \citep{scikit-learn}, Scipy \citep{2020SciPy-NMeth}, Pandas \citep{reback2020pandas}, Matplotlib \citep{Hunter:2007}. }



\appendix
\section{Software Products}\label{sec:software}

This project resulted in two open-source software packages in Python: \texttt{FLICKER} (\url{https://github.com/lyx12311/FLICKER}) and \texttt{Astraea} (\url{https://github.com/lyx12311/Astraea}).

\begin{figure}[h!]
    \centering
    \includegraphics[width=0.4\textwidth]{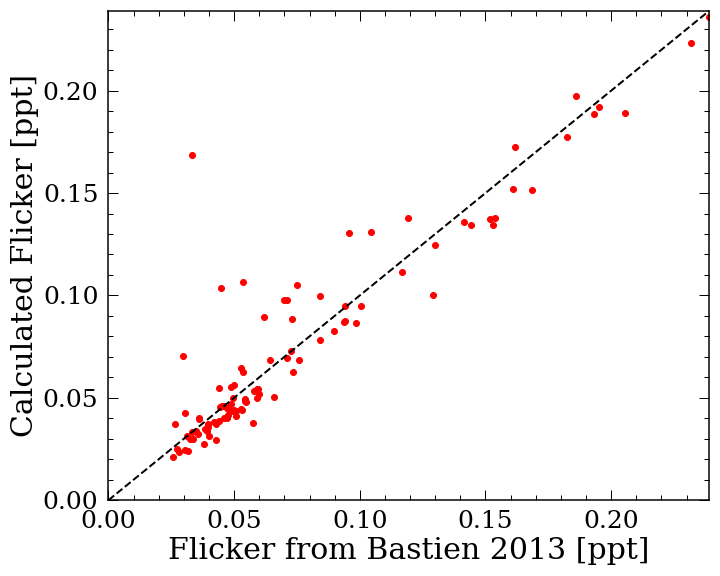}
    \caption{8-hour Flicker values calculated by \cite{Bastien2013} versus those calculated with the software package \texttt{FLICKER}. The results are consistent with one another.}
    \label{fig:FLICKER}
\end{figure}

\texttt{FLICKER} can be used to calculate flicker for one light curve or multiple light curves.
It calculates the median flicker across light curves if passed a multi-dimension array. 
Figure~\ref{fig:FLICKER} shows the comparison between flicker values provided in \cite{Bastien2013} and those calculated with \texttt{FLICKER} for 100 {\it Kepler} stars listed in their paper.

\texttt{Astraea} is a software package that includes the RF classifier and regressor trained on {\it Kepler} targets.
It can be used to recover rotation periods for any stars observed by {\it Kepler} or {\it TESS}. However, since this model is only trained on {\it Kepler} stars, any rotation periods predicted for targets outside of the {\it Kepler} field are subject to higher uncertainties. 

\bibliography{references.bib}{}
\bibliographystyle{aasjournal}



\end{document}